\documentclass[aps,prl,reprint,superscriptaddress,longbibliography]{revtex4-2}

\usepackage{amsmath}
\usepackage{xcolor}
\usepackage{textcomp}
\usepackage{graphicx} 
\usepackage{float}
\usepackage{epstopdf}
\usepackage[colorlinks,citecolor=blue,linkcolor=blue]{hyperref}

\newcommand{\SPEC}{SPEC, CEA, CNRS, Université Paris-Saclay, Gif-sur-Yvette, France}
\newcommand{\DEE}{Department of Electrical Engineering and ICT, University of Naples Federico II, Naples, Italy} 
\newcommand{\LMOPS}{LMOPS EA 4423 Laboratory, CentraleSupélec, Université de Lorraine, Metz, France}

\begin{document}

\title{Hyperchaos in a Magnetic Nanodisk Driven by Ferromagnetic Resonance}

\author{A. Kolli}
\email{amel.kolli@cea.fr}
\affiliation{\SPEC}
\author{H. Merbouche}
\affiliation{\SPEC}
\author{S. Perna}
\affiliation{\DEE}
\author{C. Serpico}
\affiliation{\DEE}
\author{D. Rontani}
\affiliation{\LMOPS}
\author{G. de Loubens}
\email{gregoire.deloubens@cea.fr}
\affiliation{\SPEC}

\begin{abstract}

We investigate the chaotic dynamics driven in the nonlinear regime of ferromagnetic resonance of an out-of-plane magnetized nanodisk in detail. By combining extensive micromagnetic simulations with time-series analysis across the control parameter space, we map the topological transitions from stable periodic orbits to strange attractors and quantify the dynamical complexity. Despite the simplicity of our nanoscale system, we evidence that it can exhibit hyperchaotic dynamics with up to three positive Lyapunov exponents in vast regions of the control plane accessible to experimental studies. Using a mode projection technique, we unveil that the generated complexity is related to the number of quantized spin-wave modes participating in the dynamics. Our findings establish magnon-spintronic nanodevices as versatile entropy sources for unconventional processing of information.

\end{abstract}

\maketitle

Chaos describes the aperiodic long-term behavior in a deterministic nonlinear system that is exponentially sensitive to initial conditions \cite{strogatz18}. It is met in a variety of dynamical systems, e.g., in electro- and opto-mechanical systems \cite{houri20,defoort21,madiot21} or in photonics \cite{soriano13,sciamanna15}, and can be leveraged for unconventional information processing, notably true random number generation (TRNG) \cite{uchida08,kanter10,madiot22}. Within this landscape, nanoscale spintronic platforms are particularly attractive as they combine strong nonlinearity, gigahertz operation and low-power requirements compatible with on-chip integration.

Ferromagnetic resonance (FMR) has been known for a long time to produce chaotic spin-wave (SW) dynamics at high power in magnetic bodies with quasi-continuous excitation spectrum \cite{nakamura82,gibson84,rezende90,wigen94}. In magnetic nanodevices, the macrospin or single-mode approximations often apply due to the strong confinement \cite{slavin09}, which for highly symmetric cases and simple excitation schemes reduces the phase space to two dimensions, thereby excluding chaos \cite{mayergoyz09}. Still, lower symmetry situations \cite{alvarez00,smith10}, modulation schemes \cite{yang07,yamaguchi19,taniguchi24a} and delayed feedback \cite{williame19,taniguchi24} allow chaotic regimes to emerge in spintronic nanodevices, as observed experimentally \cite{montoya19,kamimaki21}. Vortex-based nano-oscillators also offer a distinct route to chaotic regimes thanks to the additional degree of freedom provided by the vortex core polarity \cite{petit-watelot12,devolder19,bondarenko19,kokkinos25}.

Even in nanoscale magnetic devices, several SW modes can participate in the nonlinear dynamics \cite{mayergoyz09,bonin12,hamadeh23}. Ferromagnetic nanodisks biased with an out-of-plane field exhibit an axially symmetric ground state and a quantized SW excitation spectrum \cite{naletov11}. Despite these simplifications, they exhibit a rich variety of nonlinear dynamics when excited by an in-plane microwave field, which crucially depends on the balance between the in-plane shape anisotropy of the thin disk and the perpendicular magnetic anisotropy (PMA) of the material. When the in-plane anisotropy dominates, the strong nonlinear frequency shift leads to foldover and a large coherent precession of magnetization can be sustained by the resonant excitation field \cite{li19}. On the contrary, when the PMA compensates the shape anisotropy, the nonlinear frequency shift vanishes and the coherent precession gets destroyed at moderate amplitude by an instability caused by nonlinear interactions between the SW eigenmodes \cite{ngouagniayemeli25}. In this case, the latter provide internal degrees of freedom to the system, in which high-dimensional chaotic states can be looked for.

Hyperchaos is a higher dimensional version of deterministic chaos, featuring more than one positive Lyapunov exponent and increased dynamical complexity. It has been reported experimentally in only a few systems, including lasers with delayed feedback \cite{fischer94,deng22} and magnonic active ring resonators \cite{bir20}. In such extended systems, the high dimensionality relies on feedback, which limits physical insight into the hyperchaotic nature of the dynamics. On the contrary, a route to hyperchaos was recently proposed in a model optomechanical system, where all degrees of freedom can be accessed \cite{halef25}. Hyperchaotic dynamics has also been evidenced in numerical models of hydromagnetic convection \cite{macek14} and semiconductor superlattices \cite{mompo21}. Here, we identify hyperchaotic regimes in a magnetic nanodisk simply driven by a single-frequency microwave field. Combining micromagnetic simulations with nonlinear time-series analysis, we evidence regimes with up to three positive Lyapunov exponents and show that this hyperchaos is enabled by the proliferation of SW modes participating in the dynamics. The number of active modes and the entropy rate follow consistent trends across the control plane, linking modal content to dynamical complexity.


\begin{figure}
	\centering
        \includegraphics[width=\columnwidth]{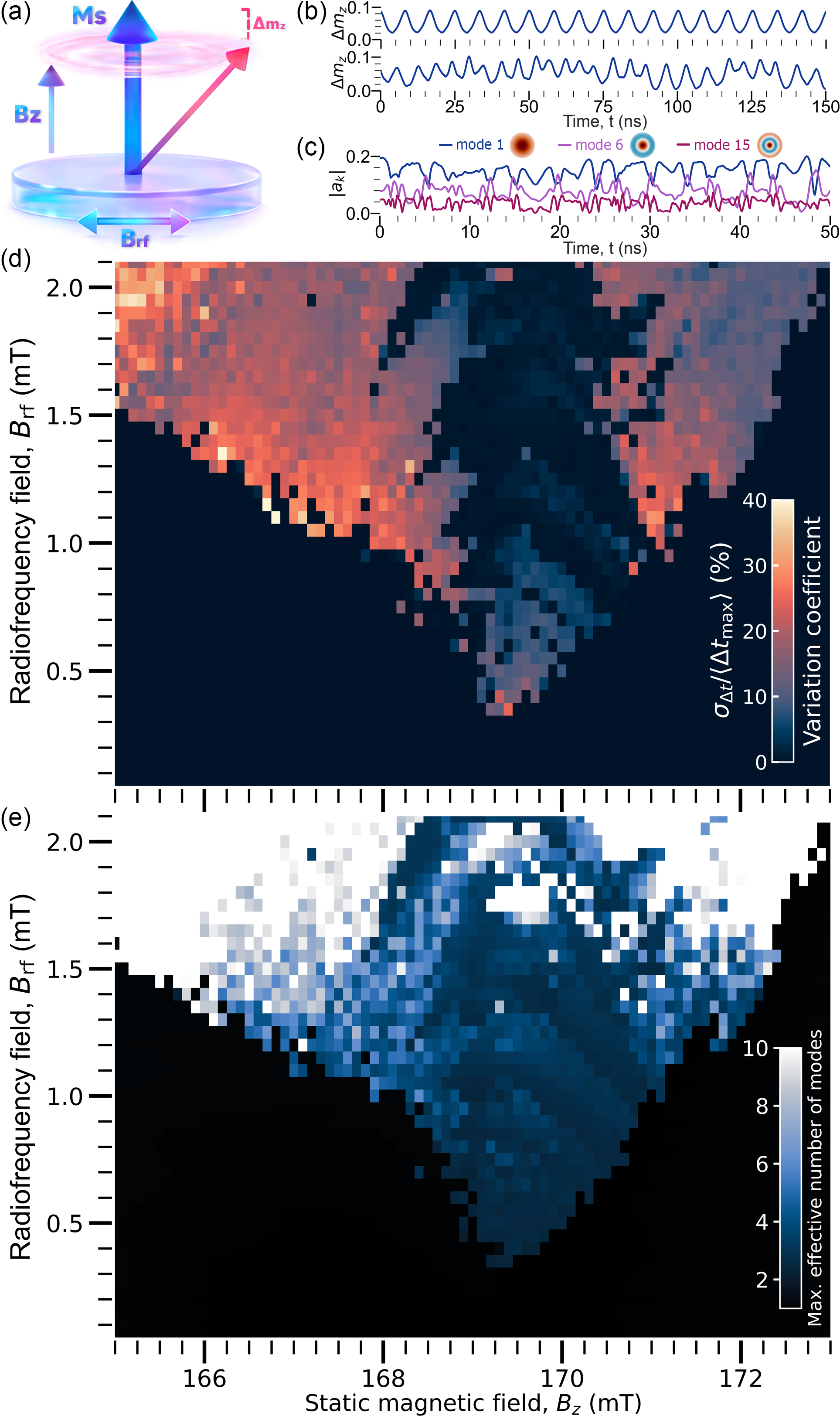}
    	\caption{  
            (a)~Magnetic nanodisk driven by an in-plane rf field $B_\mathrm{rf}$ ($f_\mathrm{rf} = 5$~GHz) under an out-of-plane static field $B_z$.
            (b)~Examples of temporal traces of $\Delta m_z$ at $B_z = 170.7$~mT with $B_\mathrm{rf} = 1.0$~mT (top) and $B_\mathrm{rf} = 2.0$~mT (bottom). 
            (c)~Corresponding temporal evolution of the amplitudes of the three main modes involved in the dynamics. 
            (d)~Variation coefficient of the peak-to-peak time intervals of $\Delta m_z$ in the $(B_z, B_\mathrm{rf})$ plane. 
            (e)~Maximum mode participation number across the same parameter space.}
	\label{fig:system_overview}
\end{figure}

The studied system consists of a 30~nm thick magnetic nanodisk with diameter 700~nm magnetized out-of-plane by a static bias field $B_z$ and driven into FMR by an in-plane linearly polarized microwave field $B_\mathrm{rf}$, as schematized in Fig.~\ref{fig:system_overview}a. Its magnetic parameters \cite{supplemental} correspond to those of a former experimental study, where a self-modulation instability was evidenced in high-power FMR \cite{ngouagniayemeli25}. In particular, its small damping parameter, $\alpha = 10^{-3}$, allows to enter the nonlinear regime of FMR at small amplitude $B_\mathrm{rf}$ of the driving field, typically well below 1~mT. Micromagnetic simulations are performed using the GPU-accelerated mumax$^3$ software \cite{vansteenkiste14}, which integrates the Landau-Lifshitz-Gilbert equation governing the dynamics. To obtain the phase diagram of the dynamics, the time-harmonic excitation ($f_\mathrm{rf} = 5$~GHz, fixed throughout this work) is turned on at $t = 0$ starting from the equilibrium configuration at each value of the control parameters $(B_z, B_\mathrm{rf})$. Cartesian components of the spatially averaged magnetization and snapshots of the full magnetization field $\mathbf{m}(\mathbf{r}, t)$ are then recorded as a function of time \cite{supplemental}. In all that follows, we denote $m_z = M_z/M_s$ the reduced longitudinal magnetization and $\Delta m_z = 1 - m_z$ its deviation from saturation.

At low rf amplitude, the steady-state dynamics corresponds to the constant-amplitude oscillation of the transverse components of magnetization at the forced frequency, resulting in a reduced longitudinal component $m_z$ stable in time. Above a certain critical amplitude of $B_\mathrm{rf}$, however, a self-modulation instability takes place, resulting in temporal oscillations of $m_z$ which never reach a stationary state, as shown in Fig.~\ref{fig:system_overview}b. This phenomenon results from the nonlinear coupling between the quantized SW modes of the magnetic disk, as demonstrated in Ref.~\cite{ngouagniayemeli25}. Below the instability threshold, only the fundamental mode of eigenfrequency close to 5~GHz is excited by the rf field and has a constant amplitude in the steady state. Above the threshold, its amplitude strongly oscillates and other eigenmodes get involved in the dynamics, as shown in Fig.~\ref{fig:system_overview}c, where a mode projection technique \cite{daquino09,massouras24} is used to evaluate the amplitudes $|a_k|$ of the three main modes as a function of time from the magnetization field snapshots.



Due to the resonance phenomenon that primarily drives the fundamental mode, the critical region where the system exhibits unstable dynamics has a typical tongue shape in the $(B_z, B_\mathrm{rf})$ control plane. Depending on the operating point in the instability region, the temporal oscillations of $m_z$ can be periodic or irregular, as can be seen in Fig.~\ref{fig:system_overview}b. To quantitatively evaluate this irregularity, we detect the local maxima of $\Delta m_z(t)$ during the last 400~ns of each simulation, extract the peak-to-peak time intervals $\Delta t_i$ and compute the variation coefficient $c_v = \sigma_{\Delta t}/\langle \Delta t \rangle$, where $\sigma_{\Delta t}$ and $\langle \Delta t \rangle$ are the standard deviation and mean of the peak-to-peak intervals, respectively. The resulting map (Fig.~\ref{fig:system_overview}d) shows the tongue-shaped instability region, with a minimal critical field of about 0.3~mT at $B_z \approx 169$~mT, close to the linear FMR resonance field at 5~GHz \cite{ngouagniayemeli25}. At this field, the variation coefficient reaches values of 10--20\% at the bottom of the tongue, and then decreases to zero and displays small oscillations as $B_\mathrm{rf}$ is increased. At lower fields ($B_z < 168$~mT) and higher fields ($B_z > 171$~mT), the variation coefficient is everywhere larger than 10\%, reaching values above 30\% just above the critical line and at sufficiently large $B_\mathrm{rf}$. These observations are a first indication that the complexity of the dynamics depends on the control parameters.

\begin{figure*}
	\centering
        \includegraphics[width=\textwidth]{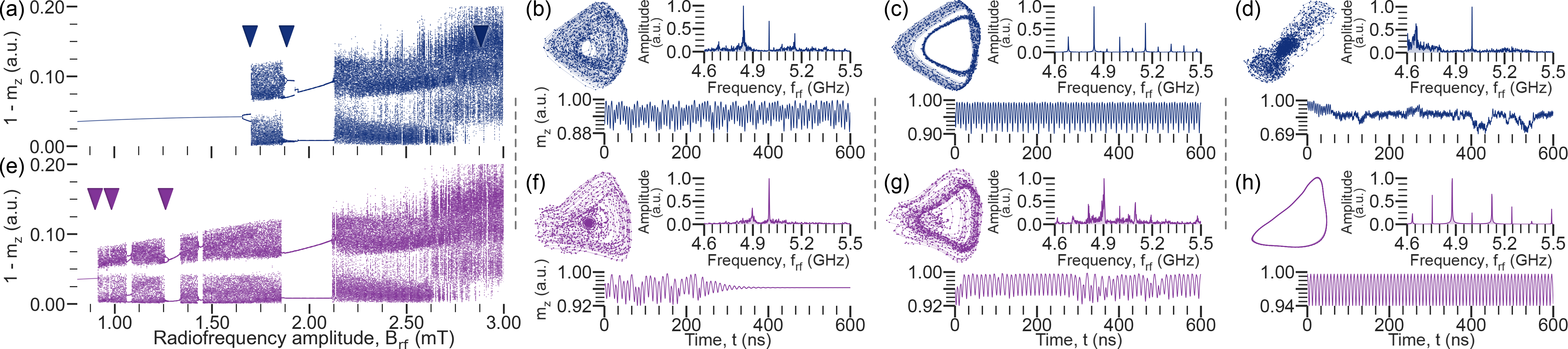}
    	\caption{
            (a)~Forward (blue) and (e)~backward (red) bifurcation diagrams constructed from the local extrema of $\Delta m_z$ at $B_z = 168.5$~mT.
            (b--d)~Three operating points during the forward sweep ($B_\mathrm{rf} = 1.745$, $1.875$ and $2.825$~mT): reconstructed phase-space attractor (top left), FFT spectrum (top right) and temporal trace (bottom).
            (f--h)~Three operating points during the backward sweep ($B_\mathrm{rf} = 0.915$, $0.940$ and $1.285$~mT), same layout.}
	\label{fig:bifurcation}
\end{figure*}

Using the mode projection technique mentioned above, we also evaluate the effective number of eigenmodes participating in the dynamics across the control plane (Fig.~\ref{fig:system_overview}e). For this, we decompose the magnetization onto the SW eigenbasis \cite{daquino09} and compute the Shannon participation number \cite{sanchezdehesa22}
\begin{equation}
    P_s = \exp\!\left(-\sum_k p_k \ln p_k\right), \quad p_k = \frac{|a_k|^2}{\sum_k |a_k|^2}
\end{equation}
which reduces to 1 when a single mode carries all the energy and grows with the number of active modes. The plotted quantity $P_s^\mathrm{max}$ is the maximum of $P_s$ over the analyzed time window. Its map mirrors the variation coefficient: below the instability boundary $P_s^\mathrm{max} < 2$, while above it at least two modes are simultaneously active. The small oscillations of $P_s^\mathrm{max}$ in the region above the tip of the tongue (values between 2 and 4) mimic those of the variation coefficient, and a proliferation of modes ($P_s^\mathrm{max} > 10$) occurs in the regions where the variation coefficient is largest. This correspondence shows that the temporal irregularity is directly linked to the multimode character of the SW dynamics.

To go beyond these scalar indicators and characterize the topology of the dynamical states, we reconstruct the phase-space trajectory of the magnetization using Takens' delay embedding \cite{takens81}, which recovers the topological structure of the full dynamical attractor from a single scalar observable. As the driving field increases, the reconstructed manifold transitions from a simple limit cycle, through a regime of intermittency, and ultimately expands into a dense, bounded strange attractor. This topological reconstruction confirms that the observed aperiodic dynamics correspond to deterministic chaos unfolding on a low-dimensional manifold, motivating the quantitative bifurcation analysis presented in the  following.


To resolve the fine structure of the dynamical transitions, we now perform quasi-static sweeps \cite{supplemental} of the rf amplitude at fixed $B_z$, in contrast with the independent simulations of Fig.~\ref{fig:system_overview}. In this scheme, the final magnetization state at each step serves as the initial condition for the next, so the system retains a dynamical memory of its history. We first focus on $B_z = 168.5$~mT, which illustrates the morphological diversity of the accessible dynamical states. The bifurcation diagram is constructed by extracting the local extrema of $\Delta m_z(t)$ at each driving amplitude. During the forward sweep (Fig.~\ref{fig:bifurcation}a), the system exhibits a single stable branch for $B_\mathrm{rf} < 1.6$~mT, characteristic of the period-1 oscillation of the uniform mode. Above this threshold, the uniform mode destabilizes through parametric excitation of secondary SW modes, and the system undergoes a cascade of bifurcations involving intermittency and period-doubling before merging into dense chaotic continua. This chaotic background is interrupted by well-defined periodic windows between 1.8 and 2.1~mT, where the dynamics temporarily collapse onto low-period limit cycles before a subsequent crisis reinstates the chaotic state.

\begin{figure*}
	\centering
        \includegraphics[width=\textwidth]{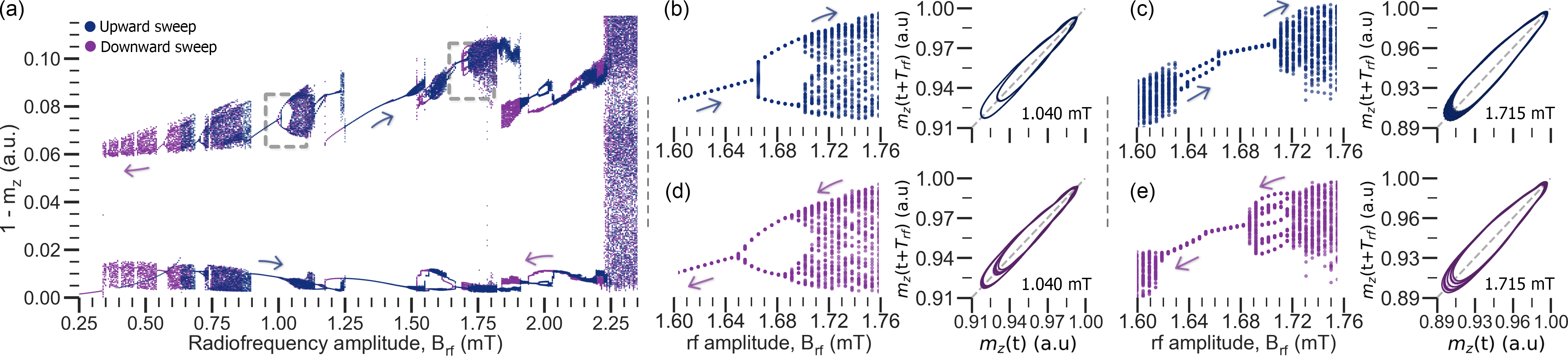}
    	\caption{
            (a)~Bifurcation diagrams for the forward (blue) and backward (red) quasi-static sweeps of $B_\mathrm{rf}$ at $B_z = 170.0$~mT.
            (b, c)~Magnified bifurcation regions and corresponding delayed return maps from the continuous time trace $m_z(t + T_\mathrm{rf})$ vs $m_z(t)$ during the forward sweep, at $B_\mathrm{rf} = 1.040$ and $1.715$~mT.
            (d, e)~Same for the backward sweep at the same operating points.}
	\label{fig:bifurcation1700}
\end{figure*}

Three operating points sampled during the forward sweep illustrate distinct dynamical regimes (Fig.~\ref{fig:bifurcation}b--d). At $B_\mathrm{rf} = 1.745$~mT, the time trace shows aperiodic amplitude modulations, the FFT spectrum develops a continuous noise floor, and the reconstructed attractor traces a dense strange attractor — the signatures of deterministic chaos. At 1.875~mT, the system exits the chaotic zone: the time trace becomes multi-periodic, the FFT presents sharp discrete peaks, and the phase-space trajectory collapses onto a limit cycle at the onset of a period-doubling bifurcation. At 2.825~mT, the dynamics reaches a strongly chaotic state with a qualitatively different attractor topology and larger amplitude excursions, consistent with an increased number of active modes and reminiscent of SW turbulence in extended systems \cite{zakharov75,bryant87,lvov94}.

The backward sweep (Fig.~\ref{fig:bifurcation}e) reveals pronounced hysteresis: the chaotic regime persists down to rf amplitudes well below the forward onset threshold. This asymmetry arises because the system, once trapped in a chaotic basin of attraction, continues to evolve within it even at lower driving powers where stable periodic orbits coexist. The chaotic state is only abandoned when the strange attractor is destroyed, typically via a boundary crisis. This path-dependent behavior indicates generalized multistability, where several coexisting basins of attraction overlap for the same parameter set.

Three points during the backward sweep further illustrate this behavior (Fig.~\ref{fig:bifurcation}f--h). At $B_\mathrm{rf} = 1.285$~mT, the dynamics has settled onto a stable period-1 limit cycle, marking the starting point of the bifurcation cascade. At 0.940~mT, the system exhibits intermittency, reflecting a direct competition between multi-periodic and chaotic regimes. At 0.915~mT, the time trace captures a transient fade-out at the end of an isolated chaotic island.

The phase-space trajectories shown in Fig.~\ref{fig:bifurcation}b--h are obtained via Takens' delay embedding of the scalar signal $\Delta m_z(t)$. For the chaotic operating points, convergence of the attractor geometry is reached for embedding dimensions between 5 and 8. Despite the complexity of the underlying multi-magnon scattering, the dynamics unfold on a low-dimensional manifold that can be tracked and classified. Such effective reductions had already been reported in early studies of high-power FMR in bulk samples, where two- or three-mode models could capture the bifurcation cascades and chaotic regimes despite the involvement of a much larger degenerate SW manifold \cite{zhang88,rezende90}. The distinct topologies observed across the control parameter, from limit cycles to folded strange attractors, motivate a closer examination of the bifurcation fine structure, which we now carry out at $B_z = 170.0$~mT.


At this static field, the forward and backward bifurcation diagrams (Fig.~\ref{fig:bifurcation1700}a) display a marked divergence in their transition thresholds. During the forward sweep, the onset of chaos is abrupt. In the backward sweep, the chaotic regimes extend significantly toward lower rf amplitudes, consistent with the boundary-crisis mechanism discussed previously. The bifurcations in the backward path also appear smoother and more progressive than their forward counterparts.

To resolve this directional asymmetry, we construct delayed return maps by plotting $m_z(t + T_\mathrm{rf})$ against $m_z(t)$ for selected operating points where both sweeps undergo a bifurcation simultaneously (Fig.~\ref{fig:bifurcation1700}b--e). At $B_\mathrm{rf} = 1.040$~mT (panels b, d), both sweeps exhibit a transition at the same rf amplitude, yet their fine structure differs: the forward sweep shows a compact, abrupt bifurcation, while the backward sweep resolves a structured period-doubling cascade with distinct concentrated bands. At $B_\mathrm{rf} = 1.715$~mT (panels c, e), the contrast is even more pronounced. The backward return map displays a comb-like sequence of branches connecting the chaotic continuum to the periodic state, characteristic of a progressive period-doubling route. The forward sweep at the same amplitude shows a less resolved transition. In all cases, the return maps share an asymmetrical geometry: the upper-right region concentrates the magnetization maxima, which remain compact across successive rf cycles, while the lower region splits into localized bands corresponding to the resolved doubling cascade.

\begin{figure}
	\centering
        \includegraphics[width=\columnwidth]{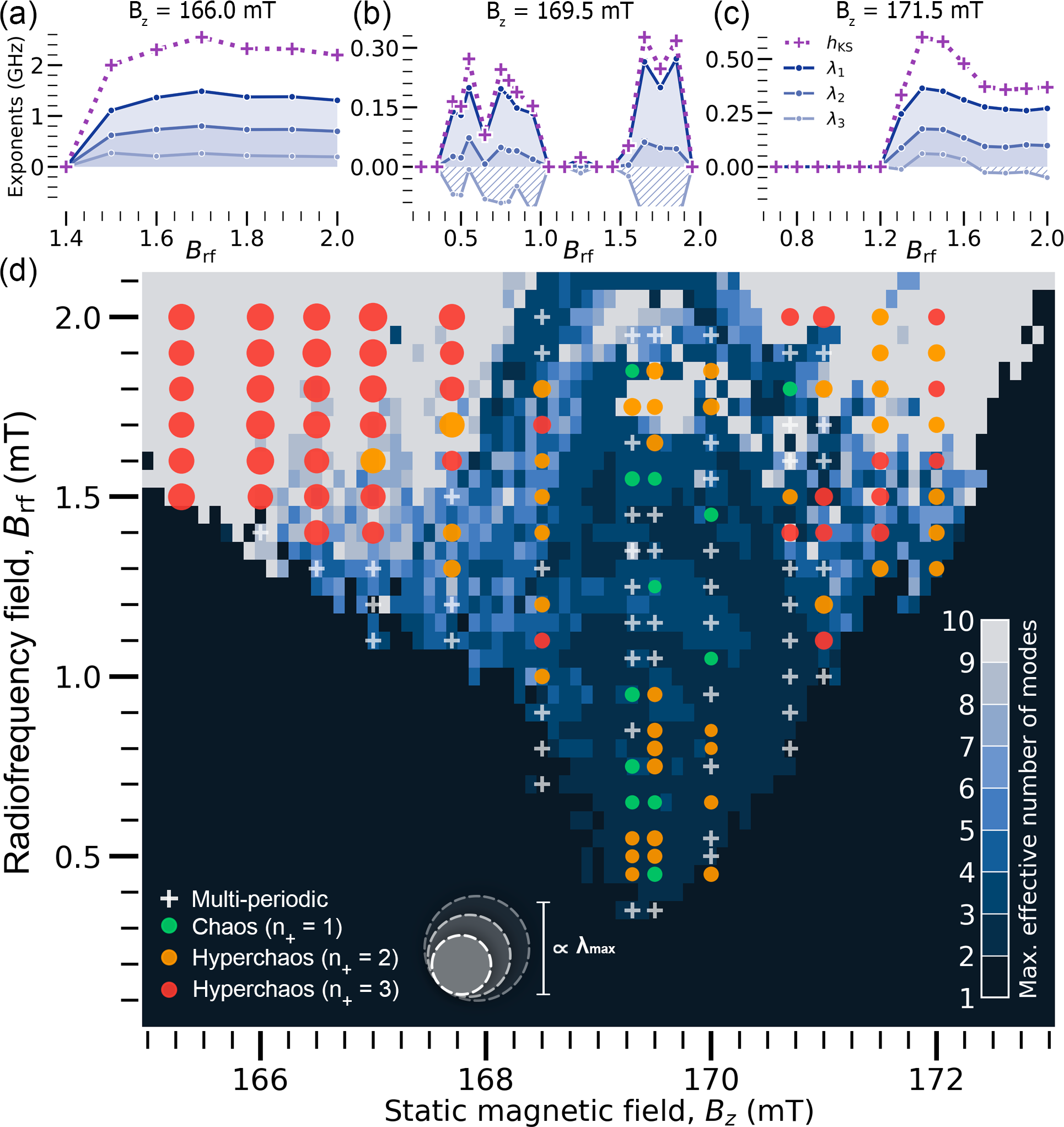}
    	\caption{(a--c)~Lyapunov spectrum ($\lambda_1$, $\lambda_2$, $\lambda_3$) and Kolmogorov-Sinai entropy $h_{KS}$ (dashed red) as a function of $B_\mathrm{rf}$ at fixed $B_z = 166.0$, $169.5$ and $171.5$~mT respectively.
            (d)~Classification map of the chaotic dynamics in the $(B_z, B_\mathrm{rf})$ plane: multi-periodic ($n_+ = 0$, crosses), chaos with $n_+ = 1$ (green), hyperchaos with $n_+ = 2$ (orange) and $n_+ = 3$ (red). The marker size is proportional to $\lambda_\mathrm{max}$. The background color recalls the maximum mode participation number (cf. Fig.~\ref{fig:system_overview}e).}
	\label{fig:lyapunov_analysis}
\end{figure}


The above topological reconstructions distinguish deterministic chaos from noise but do not quantify the dynamical complexity. To this end, we extract the complete spectrum of Lyapunov exponents $\lambda_i$, which measure the average rate of exponential divergence between nearby trajectories in phase space \cite{kantz03,eckmann85}. Since the system is periodically driven, we sample $\Delta m_z(t)$ stroboscopically at $f_\mathrm{rf}$ to project the flow onto a Poincar\'e section and compute the spectrum from the resulting discrete map using a Jacobian-based method with QR reorthonormalization \cite{bremen97,supplemental}.

We evaluate the Lyapunov spectrum as a function of $B_\mathrm{rf}$ at three fixed static fields (Fig.~\ref{fig:lyapunov_analysis}a--c). At $B_z = 166.0$~mT (panel~a), the instability threshold lies near $B_\mathrm{rf} \approx 1.4$~mT. Once crossed, three exponents become positive simultaneously, with $\lambda_1$ reaching $\sim 1$~GHz. The Kolmogorov-Sinai entropy rate, estimated as $h_{KS} \approx \sum_{\lambda_i > 0} \lambda_i$, exceeds 2~GHz. The system enters sustained hyperchaos immediately above the threshold. At $B_z = 169.5$~mT (panel~b), the threshold drops to $B_\mathrm{rf} \approx 0.3$~mT, reflecting the resonant character of the FMR-driven dynamics. The spectrum displays a finer structure: chaotic windows with two positive exponents alternate with periodic intervals where all exponents return to zero, consistent with the periodic windows observed in the bifurcation diagrams. The exponent magnitudes remain moderate ($\lambda_1 \sim 0.1$--$0.3$~GHz). At $B_z = 171.5$~mT (panel~c), two, then three exponents become positive as $B_\mathrm{rf}$ increases above the threshold value $\approx 1.2$~mT, before one of the three largest exponents turns negative by further increasing $B_\mathrm{rf}$. The entropy rate is maximal in the rf window where three positive Lyapunov exponents are extracted, as could be expected.

The Kaplan-Yorke dimension $D_{KY}$ measures the geometric extent of the attractor, while the correlation dimension $D_2$ from a Grassberger-Procaccia analysis \cite{kantz03,supplemental} probes how uniformly the trajectory visits it. Their difference, the multifractal gap $\Delta = D_{KY} - D_2$, quantifies the non-uniformity of the invariant measure. At $B_z = 169.5$~mT, a single rf increment from 1.55 to 1.65~mT spans the chaos-to-hyperchaos transition: $n_+$ jumps from 1 to 2 and $h_{KS}$ rises 
sixfold (0.05 to 0.33~GHz), yet $D_{KY}$ is conserved within 1\% ($3.44 \to 3.42$). Meanwhile, $D_2$ drops from $1.38 \pm 0.04$ to $1.19 \pm 0.01$, widening $\Delta$ by 8\%. The transition is thus a change of dynamical content on a geometrically conserved attractor: unstable directions double and information generation rises sixfold, while the $D_2$ drop at conserved $D_{KY}$ signals a more concentrated invariant measure on the same geometric support.

To show that the generation of high entropy is a tunable property of the system, we map the number of positive Lyapunov exponents $n_+$ across the full $(B_z, B_\mathrm{rf})$ control plane (Fig.~\ref{fig:lyapunov_analysis}d). Each point is classified as multi-periodic ($n_+ = 0$, crosses), chaotic ($n_+ = 1$, green), or hyperchaotic ($n_+ = 2$, orange; $n_+ = 3$, red), with the marker size proportional to the largest exponent $\lambda_\mathrm{max}$. The background color scale reproduces the mode participation number from Fig.~\ref{fig:system_overview}e. The resulting classification mirrors the tongue instability boundary. On the left flank ($B_z < 168$~mT), sustained hyperchaos with three positive exponents and large $\lambda_\mathrm{max}$ dominates at high rf amplitudes. Near the tip of the tongue ($B_z \approx 169$~mT), the onset of chaos occurs at lower rf thresholds, but the dynamics alternate between chaotic and periodic states. On the right flank ($B_z > 170$~mT), hyperchaos persists but with smaller exponents, consistent with the weaker entropy rates observed in the spectral cuts. Ordinary chaos with a single positive exponent (green markers) is sparse and confined to the boundaries between periodic and hyperchaotic regions \cite{halef25}. Moreover, the overall correlation between $n_+$ and $P_s^\mathrm{max}$ confirms that multimode activation is a prerequisite for hyperchaos. Detailed inspection of the generated SW modes, in which the nonlinear interactions at play establish a hierarchy \cite{bonin12}, could provide additional insight into the classification of the chaotic regimes.

In sum, we have shown that a magnetic nanodisk driven into FMR by a time-harmonic field can exhibit hyperchaos with up to three positive Lyapunov exponents. This high-dimensional chaos is enabled by the nonlinear coupling between its quantized SW eigenmodes and remains confined to a finite-dimensional strange attractor, as confirmed by the Kaplan-Yorke dimension. The entropy rate reaches $h_{KS} \approx 2.5$~GHz in the most chaotic regions ($\sim3.6$~Gbit/s of intrinsic information generation), positioning such nanodevices as tunable entropy sources for TRNG. The exploited nonlinear SW dynamics relies on a near cancellation of the effective anisotropy, also achievable in magnetic tunnel junctions by tuning their PMA \cite{devolder19a}. In such devices, the chaotic signal would be directly readable via magnetoresistance, opening a route toward integrated nanoscale TRNG, which has already been demonstrated using thermal stochasticity \cite{phan24,sidielvalli25}. More broadly, the well-defined and hysteretic boundaries between periodic, chaotic and hyperchaotic states in the $(B_z, B_\mathrm{rf})$ phase diagram are relevant to the control and synchronization of chaos for secure communications \cite{ott90,peterman95,argyris05}, as well as to unconventional computing schemes exploiting chaotic dynamics and multistability \cite{langton90,sinha98,yoo20,liu25,nishimura26}. Finally, our results suggest that turbulent regimes of SW dynamics, long studied in extended systems \cite{zakharov75,bryant87,lvov94}, can also be sustained in a single nanostructure.

\textit{Acknowledgments—} This work is supported by France 2030 government investment plan managed by the French National Research Agency under grant reference PEPR SPIN -- SpinCom ANR-22-EXSP-0005.

\textit{Data availability—} The data that support the findings of this Letter are openly available [zenodo].

\bibliography{chaos}

\end{document}